\documentstyle[psfig]{l-aa}
\def\A{\nabla_\lambda(\sigma_f)}
\def\B{\sigma_{f,2000}}
\def\BR{\sigma_{f,1250}^{\mathrm{rest}}}
\def\dd{{\mathrm{d}}}
\def\sigf{\sigma_f(\lambda)}
\def\SF{Seyfert-like objects}
\def\Var{{\mathrm{Var}}}
\def\BN{\overline{N}}
\def\BL{\overline{L}}

\begin{document}
\renewcommand{\bottomfraction}{0.5}
\thesaurus{3(13.21.1; 11.01.2; 11.17.3; 11.02.1)}
\title{Rest-frame variability of quasars and Seyfert galaxies in the ultraviolet: Constraints on the discrete-event models}
\author{S.\ Paltani \inst{1} \and T. J.-L.\ Courvoisier \inst{2,3}}
\institute{CESR, 9, av.\ du Colonel-Roche, 31028 Toulouse cedex, France
\and
INTEGRAL Science Data Centre, ch.\ d'Ecogia 16, CH-1290 Versoix, Switzerland
\and
Geneva Observatory, ch. des Maillettes 51, CH-1290 Sauverny, Switzerland}
\offprints{Thierry J.-L. Courvoisier (ISDC); e-mail: Thierry.Courvoisier@obs.unige.ch}
\date{Received 6 September 1996 / Accepted 13 January 1997}
\maketitle
\markboth{S.Paltani \& T.J.-L.Courvoisier: Rest-frame variability of AGN and discrete-event models}{S.Paltani \& T.J.-L. Courvoisier: Rest-frame variability of AGN and discrete-event models}

\begin{abstract}
A study of the ultraviolet continuum variability (Paltani \& Courvoisier \cite{PC94}) has shown that the relative variability of quasars and Seyfert galaxies decreases when the luminosity increases. The spectral information included in the IUE spectra allows us to study this dependence in the rest frame of the objects. The trend is strengthened by the general property that active galactic nuclei vary more at short wavelengths than at long wavelengths in the ultraviolet domain. The scatter observed in all other studies is still present. An important part of this scatter may however be explained if one tries to estimate the uncertainties on the variability due to the sampling.

We discuss the variability using the concept of discrete events. The trend between variability and luminosity is described by a power-law with an index $-0.08$, which is incompatible with the power-law of index $-1/2$ predicted by the most general discrete-event models in which the change in average luminosity is due to differences in average event rates exclusively. Several biases are investigated, but we conclude that the $-1/2$ index is definitely inconsistent with the data. A flat relationship is however possible, if some bias has been underestimated.

We propose different ways whereby discrete events may produce a different variability--luminosity relationship: by changing the luminosity or the life time of the events, or by introducing interdependence between the events. The latter possibility cannot produce a satisfactory relationship. Using the former possibilities, we do not find any ``natural'' explanation for the variability--luminosity relationship in the context of discrete-event models. This is possibly an indication that explanations in which variability is not expressed in terms of discrete events should be favoured.


\end{abstract}

\section{Introduction}
The statistical properties of the ultraviolet variability of active galactic nuclei (AGN) have been investigated by Paltani \& Courvoisier (\cite{PC94}) (hereafter PC94). This study, which was based on spectra obtained by the {\em International Ultraviolet Explorer} (IUE) from the ULDA ({\em Uniform Low Dispersion Archive}) database, has been done in the observer's frame. In addition to the variability at 2\,000 \AA, we quantified also the change of variability with the wavelength, which showed that all classes of AGN have a larger variability at short wavelength than at long wavelength in the IUE domain. In this paper we shall consider only the objects that have been called \SF\ in PC94, i.e.\ Seyfert 1 galaxies, radio-quiet quasars, and low-polarization radio-loud quasars.

The relationship between variability and luminosity in quasars and Seyfert galaxies has been investigated by many authors. Most of them found an anti-correlation between these two quantities (Pica \& Smith \cite{PS83}; Cristiani et al.\ \cite{Cal90}; PC94; Hook et al.\ \cite{Hal94}), while others found a correlation or no correlation at all (Bonoli et al.\ \cite{Bal79}; Tr\`evese et al.\ \cite{Tal89}; Giallongo et al.\ \cite{Gal91}). Rediscussion of these last three papers by Hook et al.\ (\cite{Hal94}) seems to show that all the samples are compatible with the existence of an anti-correlation.

All these studies have used estimates of the variability in the observer's frame, as they can be much more easily obtained than those in the objects' rest frame. Therefore the correct relationship between variability and luminosity is still unknown. This relationship is very important, as it can be compared with the predictions of some AGN models, in particular the starburst model of Terlevich (\cite{T92}). In this paper we establish this relationship in the objects' rest frame, and we discuss some biases that can affect it. We then check the compatibility between our result and the predictions of these models.

We formalize in the most general way the concept of ``discrete-event models'', i.e.\ models where the total light curve is the sum of a series of light curves, each one associated with one ``event'', whatever its physical nature, and we compare their predictions with the above constraint. We give the general relationship between variability and luminosity for these models. We study two different ways by which this relationship can be modified. In the first one we assume that some of the event parameters may actually depend on the luminosity. In the second one we examine the effects of the introduction of interdependence between the events.

\section{Rest-frame variability in the ultraviolet}
\subsection{Determination of the rest-frame variability}
In this paper, as in PC94, we use the IUE spectra from the ULDA version 4.0 database. This database contains nearly every low-dispersion spectrum obtained by the IUE satellite before 1992. The period covered by the ULDA archive is therefore about 14 years. The wavelength range covered by these spectra, combining short- and long-wavelength cameras, is from 1200 \AA\ to 3200 \AA. The large number of repeated observations and the spectral coverage of the IUE satellite make this satellite ideal for the determination of the AGN variability properties.

In PC94 we estimated for each object the wavelength-dependent variability, $\sigf$. In all this paper, as in PC94, ``variability'' should be understood as the standard deviation of the flux divided by the mean flux. For each object we approximated $\sigf$ by a linear relationship:
\begin{equation}
\sigf\simeq \A\cdot\frac{\lambda - 2000~{\mathrm\AA}}{1000~{\mathrm\AA}}+\B
\end{equation}
$\B$ is expressed in percents of the mean flux and $\A$, the slope of $\sigf$, is expressed in percents of the mean flux per 1\,000 \AA\ (cf.\ Fig.~1 of PC94). The values of $\B$ and $\A$ can be found in the Appendix of PC94. The definition of these parameters has only a descriptive purpose and does not have any predictive value outside the wavelength range studied. Using these two parameters, we can derive the AGN variability at a given wavelength, 1\,250 \AA\ in our case, in the rest frame of the objects:
\begin{equation}
\BR\simeq \B+\A\cdot\frac{1\,250\cdot(1+z)-2\,000}{1\,000}
\end{equation}
We feel that it is important not to extrapolate the variability outside the IUE wavelength range. We therefore consider in this paper only the objects that have a redshift smaller than 1.3.

\subsection{Uncertainties on the rest-frame variability}
It is important to estimate the uncertainty on $\BR$, as we shall be concerned in this paper with the relationship between variability and another parameter, namely the luminosity. The uncertainty on $\BR$ has two origins. First it is due to the propagation of the uncertainties on $z$, $\B$, and $\A$ from PC94. As explained in PC94, the two last uncertainties are the uncertainties on the least-squares linear regression and do not take into account the effect of the sampling. A second term must therefore be included to allow for the effect of the sampling. We propose here a method that is an improvement of the discussion of Sect.~3.2 in PC94.

The first step is the simulation of a light curve that has variability  properties comparable to those of our objects. We assume that the AGN light curves are characterized by a broad Fourier power spectrum which follows a power-law with an index between $-1$ (flicker noise) and $-2$ (random walk). These has been verified for at least two \SF: 3C 273 (Kunkel \cite{K72}) and NGC 4151 (Longo et al.\ \cite{Lal96}) (see also Press \cite{P78}). An example of a light curve simulated by a ``random walk'' can be seen in Fig.~1 of PC94. We rescale these simulated light curves to obtain a variability comparable to the ones observed in our objects. Then we draw $n$ epochs from an uniform distribution on the period covered by ULDA and compare the variability estimated with these $n$ points with the variability of the simulated light curve. Repeating this last step many times, we can estimate the uncertainty produced by the sampling from the dispersion of the estimates of the variability.

As the indices are smaller or equal to -1, a low-frequency cut-off has to be chosen if one wants to avoid infinite variability. On the basis of archival data of 3C 273 (Angione \& Smith \cite{AS85}), it appears that variability on time scales longer than 20 years is not dominant. If this is valid for all AGN, a 14-year period should contain almost all the power of the Fourier spectrum. The effect of variations on longer time scales and of time dilation in high-redshift objects are discussed in Sect.~\ref{verylong} and \ref{tim-dil} respectively. 

Taking 3 values for the spectral index of the power spectrum ($-1$, $-1.5$, and $-2$), 3 values for the low-frequency cut-off ((5 yr)$^{-1}$, (10 yr)$^{-1}$, and (20 yr)$^{-1}$), and 3 values for the relative variability (0.05, 0.1 and 0.5), we parametrize the relative uncertainty on $\sigma$, $\frac{\Delta\sigma}{\sigma}$, as a function of $n$: $\frac{\Delta\sigma}{\sigma}(n)=\sigma_{20}\cdot (n/20)^s$ (where $\sigma_{20}$ is the uncertainty for $n=20$), which produces good approximations. We normalize $\frac{\Delta\sigma}{\sigma}$ at $n=20$ because this value is central and is little affected by small deviations of $s$. It appears from the simulations that none of the 3 parameters modifies significantly the value of $s$, and $s=-0.56$ reproduces quite well the decrease of $\frac{\Delta\sigma}{\sigma}$, at least for $n\le 50$. On the other hand, $\sigma_{20}$ depends on the 3 parameters, but the differences are only a few percents. $\sigma_{20}$ increases when the index increases, the low-frequency cut-off decreases, or the variability increases. We therefore use the following empirical relation for the uncertainty on the relative variability due to the sampling in the worst case, i.e.\ low-frequency cut-off$=$(20 yr)$^{-1}$, and index$=-1$ (flicker noise):
\begin{equation}
\frac{\Delta\sigma}{\sigma}(n)=(0.17+2\cdot\sigma^2)\cdot \left(\frac{n}{20}\right)^{-0.56}
\end{equation}
The best case makes $\sigma_{20}$ become $(0.14+2\cdot\sigma^2)$. In practice, as the number of spectra is usually different in the short- and long-wavelength ranges, $n$ will be the larger of the two numbers. For the few cases where the number of spectra is larger than 50, we use $\frac{\Delta\sigma}{\sigma}=0.05$. It is important to note that the estimates of the variability follow rather well a Gaussian distribution as soon as $n$ is larger or equal to 4 (the distribution is already marginally acceptably Gaussian for $n=3$). Thus we can estimate the total uncertainty on $\BR$ by simply adding the variances of the two sources of dispersion.

\subsection{Correlation between variability and luminosity\label{covarlum}}
\begin{figure}[t]
\centerline{\mbox{\psfig{file=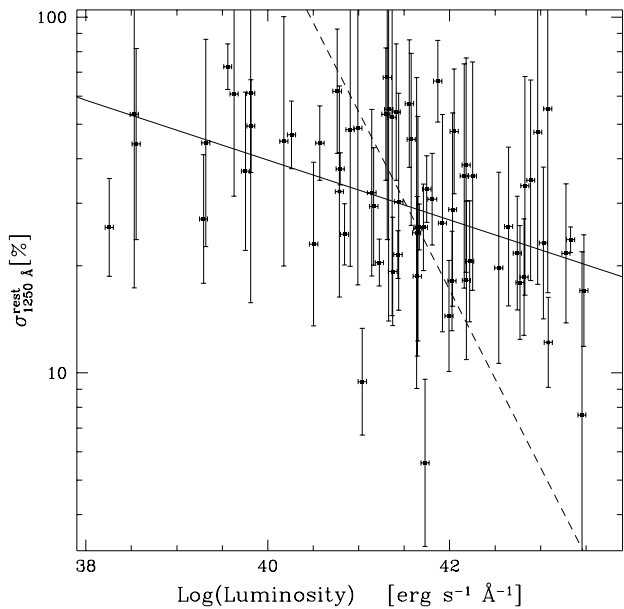,width=8.8cm}}}
\caption{\label{Blum}$\BR$ vs luminosity for \SF. The solid line is the  power-law obtained with the BCES method, and has a slope of -0.078. The dashed line is the best $\chi^2$ fit with a power-law of fixed index $-1/2$.} 
\end{figure}
Figure \ref{Blum} shows the variability $\BR$ as a function of the luminosity for all the objects classified as \SF\ in PC94 (the objects and their properties are given in the appendix of PC94). The Spearman's correlation coefficient is equal to -0.45. The probability to get such a high correlation coefficient for 68 couples drawn from two uncorrelated parent populations is about $10^{-4}$. We note that Spearman's test is valid even if the variables have uncertainties or intrinsic scatter. A real correlation may be destroyed by the dispersion, but the probability that the dispersion generates a spurious correlation is very low. Therefore, with better measurements, the correlation coefficient should in principle increase (in absolute value). It appears therefore unquestionable that a trend exists, and this trend is in the sense that the variability decreases when the luminosity increases.

To quantify this trend, one must choose an analytical formulation. Because the very large dynamic of the $x$ axis ($\sim 10^5$), the most natural one is a power-law relation in the form:
\begin{equation}
\label{vlrel}
\BR\sim \BL^{\;\eta}_{\mathrm{1250 \AA\ (rest)}},
\end{equation}
where $\BL_{\mathrm{1250 \AA (rest)}}$ is the average luminosity density of the object at 1250 \AA\ in the rest-frame of the object. The index $\eta$ of the trend can in principle be estimated. However, because of the important uncertainties, one must be careful about the selection of the method. As we determined both $x$- and $y$-axis uncertainties, we used a $\chi^2$ method  that takes into account the uncertainties on both coordinates (Press et al.\ \cite{PVTF92}). This method is valid only for linear (affine) relationships; therefore we fit the logarithm of the values. Unfortunately, we have to estimate the variances of the logarithms of the distributions, which means that the uncertainties are not Gaussian anymore. We nevertheless assume that they follow closely enough Gaussian distributions in order to apply the $\chi^2$ test. We obtain $\eta=0.084$. The probability that a linear relationship really holds is of the order of $10^{-3}$ ($\chi^2\simeq 104$ with 66 degrees of freedom). As our uncertainties are probably overestimated, some intrinsic scatter around this relationship does exist, but a large part of the dispersion can be accounted for by the uncertainties. The linear function with a slope fixed to $-1/2$ that minimize the $\chi^2$ has been plotted on Fig.\ \ref{Blum}. It is clearly unadapted to the data (its $\chi^2$ is larger than 1000). The value of the slope obtained using the usual least-squares method (which neglects the uncertainties on both axes) is $-0.078$, which is very close. 

The scatter makes the uncertainty on the slope of the relationship meaningless. Fortunately a new method has appeared that is not only applicable when uncertainties on both axes exist, but also when there is some intrinsic scatter. This method, called BCES (Akritas \& Bershady \cite{AB96}), gives a slope $\eta=-0.078\pm 0.16$. The agreement between all these methods reinforces the confidence in the value, and we use in the following:
\begin{equation}
\eta=-0.08
\end{equation}

The BCES method is, as the authors claim, very conservative. Although $\eta=0$ is formally possible, Spearman's test, which is much more efficient in detecting the existence of a slope, clearly rules out this hypothesis.

It is interesting to note that Hook et al.\ (\cite{Hal94}) have found indices of $-0.06$ (their model A) and $-0.13$ (their model B) for the same relationship, which are both compatible with our result (but see Sect.~\ref{vrrel}). Their quasar sample extends far beyond our sample in redshift and in luminosity and does not comprise any Seyfert galaxy. Moreover, as we discuss in the next section, they do not allow for several biases.

\section{Biases in the variability--luminosity relationship\label{bias}}
We have seen in Sect.~\ref{covarlum} that the slope of the variability--luminosity relationship, although statistically different from 0, is small. Therefore, we can ask whether this trend could be due to a bias. On the other hand, it is also important to ask whether the slope of the relationship might be lessened by systematic effects.

\subsection{Very-long-term variability\label{verylong}}
The ULDA 4.0 data span about 14 years. It means that variability on time scales longer than $\sim 20$ years will not contribute (or only partly) to our estimation of $\sigf$. If only low-luminosity objects vary on very long time scales, the observed $\sigma(\BL)$ relationship would be flatter than the intrinsic one. On the other hand, the decrease of variability with luminosity may be introduced by bias if only high-luminosity objects vary on very long time scales.

Very-long-term variability can have two manifestations: either slow, steady variations or short, intense flares occuring with a rate of around 1 per century, or less. In any case, this very-long-term variability implies that the Fourier power spectrum of the AGN light curves should contain a non-negligible amount of power at frequencies smaller than $(20{\mathrm{~yr}})^{-1}$. We can estimate this power for two limiting cases {\em assuming} a given intrinsic $\sigma(\BL)$ relationship:
\begin{itemize}
\item The variability is independent of the luminosity. In this case the more luminous the object, the larger the unobserved proportion of power in the Fourier spectrum at very small frequencies. Assuming that only variability on short time scales exists in low-luminosity objects, it implies that short-time-scale variability represents only 40 \% in high-luminosity objects. The variability of 3C 273 from optical archival data since 1887 (Angione \& Smith \cite{AS85}) is about 16 \%. Its variability on a 10-year basis is 12 \% (work in progress). In the Fourier spectra presented by Angione \& Smith (\cite{AS85}), much power seems to be contained between (20 yr)$^{-1}$ and (10 yr)$^{-1}$, which could easily explain the difference, but the increase in variability is much smaller than the required factor 2.5. The power contained between (100 yr)$^{-1}$ and (20 yr)$^{-1}$ seems negligible. Even though it is {\em a priori} possible that still much power exists at frequencies below (100 yr)$^{-1}$, the small increase of power between (100 yr)$^{-1}$ and (20 yr)$^{-1}$, together with the necessity of the disappearance of this power in low-luminosity objects, makes this possibility rather improbable.
\item The intrinsic $\sigma(\BL)$ relationship has a $-1/2$ slope, which is the slope predicted by many discrete-event models (see Sect.~\ref{Discrete}). In this case the variability of low-luminosity objects must have been underestimated by a factor about 30. Assuming that the variability of high-luminosity objects is correct, which means that only variability on short time scales exists in these objects, the luminosity of low-luminosity objects should be dominated in average by very-long-term variations. As short-term variability is not much smaller than 100 \% in most low-luminosity objects, it would imply that all these objects are in phases where the very-long-term variations have a very small luminosity. This is statistically impossible if the very-long-term variations are slow and steady. Short intense flares can also be dismissed, as a sufficient variability can be attained only if the average luminosity is dominated by the very rare flares by a large factor. Such flares have never been observed. Taking into account the intense monitoring of several Seyfert 1 galaxies, the hypothetical flaring state can be active only for a very small fraction of the time. Consequently, the flare intensity should be extremely large, should have been detected in the same way as supernovae are detected.
\end{itemize}
In conclusion, although we cannot completely exclude any important effect on the variability--luminosity relationship, it appears more probable that such biases are marginal. But if the variability--luminosity relationship is affected, it is certainly in the sense that the trend is artificially produced rather than in the sense that the $-1/2$ relationship is flattened.

\subsection{Time-dilation effect\label{tim-dil}}

If light curves of AGN are stochastic time series that have Fourier power spectra well described by a power-law with indices between $-1$ and $-2$, the total variability of the objects increases with the length of the sampling until this length becomes comparable to the inverse of the frequency of the low-frequency cut-off. Because of the time dilation, high-redshift objects are potentially observed for a smaller time than low-redshift ones. Let us assume that the ``standard'' index is $-2$, which corresponds to a ``random walk''. If $x(t)$ is such a time series, its structure function $S_x(\tau)$ (Rutman \cite{R78}; Simonetti et al.\ \cite{Sal85}) is linear: $S_x(\tau)\sim\tau$. Larger values of the power-spectrum index make $S_x(\tau)$ increase more slowly.

The value of the structure function is also related to the variance of the light curve:
\begin{equation}
S_x(\tau_0)\simeq 2\cdot\Var(x_{\tau_0})
\end{equation}
where $\Var(x_{\tau_0})$ is the variance of a portion of the time series $x$, whose length is $\tau_0$. However, because of the redshift of the sources, one does not measure the correct length of the time series $\tau_0$, but the observer's length, which is $\tau=\tau_0\cdot(1+z)$, where $z$ is the redshift. Therefore, assuming that the observer's lengthes of the time series are all identical, one has to modify the observed variability:
\begin{equation}
2\,\Var(x_\tau)\simeq S_x(\tau)=(1+z)\,S_x(\tau_0)\simeq 2\,(1+z)\, \Var(x_{\tau_0})
\end{equation}
which gives:
\begin{equation}
\sigma_{x,\tau}\simeq\sigma_{x,\tau_0}\cdot\sqrt{1+z}
\end{equation}

This correction is an upper-limit, because it assumes that the power spectrum extends down to frequencies much smaller than (14 yr)$^{-1}$. If one applies this correction to our data, the anticorrelation still holds with a confidence of 99.8 \%, and the slope obtained by the BCES method is $-0.058\pm 0.15$. Therefore, it seems that the decrease of the variability with the luminosity cannot be entirely explained by this effect. 

\subsection{Existence of a correlation between variability and redshift\label{vrrel}}
Luminosity and redshift are always very well correlated in flux-limited samples, since only the most luminous objects can be observed at large redshift. This leads to a correlation between variability and redshift. The question that arises is to determine whether this correlation is entirely due to the bias, or whether there also exists an intrinsic correlation. Hook et al.\ (\cite{Hal94}), using subsamples with small luminosity range, have concluded that the first alternative is the correct one.

Two effects can produce a correlation between variability and redshift. The first is evolution; unfortunately, we are much too far from a clear picture of how AGN work to guess anything about its effect, if any, on the variability--redshift correlation. The second effect is much clearer; in PC94, we showed that the variability increases with the frequency. As the observations are usually performed at a wavelength fixed in the observer's frame, this should naturally lead to a positive correlation between variability and redshift. Di Clemente et al.\ (\cite{DCal96}) found that this effect alone can account for the positive correlation between variability and redshift, which means that there is no need to invoke evolution effects.

To allow for this effect, Cid Fernandes et al.\ (\cite{CFal96}) have fitted the data of Hook et al.\ (\cite{Hal94}) with a function of two variables, the luminosity and the redshift, and three parameters, $a$, $b$, and $c$:
\begin{equation}
\log\sigma=a\cdot\log \overline{L_B}+b\cdot\log(1+z)+c,
\end{equation} 
where $\overline{L_B}$ is the mean luminosity in the optical B band and $z$ is the redshift. The parameter $a$ is is equivalent to our parameter $\eta$; $b$ takes into account the increase of variability with the frequency; and $c$ is a normalization parameter. Although evolution effects could in principle affect parameter $b$, this is not considered by Cid Fernandes et al.\ (\cite{CFal96}), because their preferred model makes use of independent events, identical for all objects (see Sect.~\ref{Discrete}).

The results of their study are: $-0.6<a<-0.2$ and $0.1<b<1.0$ (in one particular case; however other cases are qualitatively identical). It is in clear contradiction with the results cited above, in particular ours. The explanation of the difference between this result and the one of, for instance, Hook et al.\ (\cite{Hal94}) lies essentially in two points:
\begin{enumerate}
\item Cid Fernandes et al.\ (\cite{CFal96}) take into account the fact that AGN vary more at shorter wavelengths. Hook et al.\ (\cite{Hal94}) neglect entirely this effect after checking that it is not important in their sample. 
\item Cid Fernandes et al.\ (\cite{CFal96}) estimate the uncertainties on the luminosity, and take them into account in the fits. La Franca et al.\ (\cite{LFal95}) have shown that the slopes are biased towards a smaller value when the errors on the $x$ axis are discarded. 
\end{enumerate}
Since our result is very close from the one of Hook et al.\ (\cite{Hal94}), who neglect these two biases, further discussion is required.

First, we note that the increase of the $b$ parameter has an effect very close to the decrease of $a$. This implies that solutions with $a\simeq -0.5$ are indeed allowed by Cid Fernandes et al.\ (\cite{CFal96}), but it requires large values of $b$ ($\simeq 1$). Although such values may be allowed in some cases, it cannot be guaranteed that most objects follow this behaviour. The determination of the rest-frame variability removes the difficulty encountered by Cid Fernandes et al.\ (\cite{CFal96}) about the value of $b$. In our case, no hypothesis, no extrapolation is needed. As evolution is probably not a problem (the more so as all our objects have redshift smaller than 1.3), no correlation between variability and redshift is expected, apart, perhaps, from the time-dilation effect discussed in Sect.~\ref{tim-dil}.

We can object to the second point that we also took into account the uncertainties on the $x$ axis to obtain the slope. On the other hand Cid Fernandes et al.\ (\cite{CFal96}) did not estimate the uncertainties on the variability, which, in our case, are much larger than those on the luminosity. They obtain steep values for the variability--luminosity relationship only in the cases where the uncertainty on the luminosity is very large (which is, by the way, an excellent illustration of the bias described in La Franca et al.\ (\cite{LFal95})). Moreover, only the BCES method takes correctly into account the existence of intrinsic scatter in the data. Once again, we took advantage of the spectral capabilities of IUE: while Cid Fernandes et al.\ (\cite{CFal96}) have to derive $k$-correction based on unknown spectral indices, our luminosities are automatically calculated at 1250 \AA\ in the rest frame of the objects, again without any hypothesis or extrapolation. We note that the $-1/2$ slope is compatible with our data at the 1 $\sigma$ level if the uncertainties on the luminosity are multiplied by a factor 18 ($\eta=-0.14\pm 0.36$), which is clearly an overestimate (moreover, the BCES method has very conservative uncertainties).

As a consequence we infer that the difference in the result is due mostly to the assumptions that they had to make to correct from the wavelength dependence of the variability and to obtain rest-frame luminosities.

\section{Variability from discrete events\label{Discrete}}
\subsection{The concept of the discrete-event models\label{proc_Poiss}}
It is  very convenient to think of the variability as due to the superimposition of perturbations on a constant source. The luminosity density at frequency $\nu$ can then be written:
\begin{equation}
\label{eq-dis_ev}
L_\nu(t)=C_\nu+\sum_i~e_{\nu,i}(t-t_i),
\end{equation}
where $C_\nu$ is the constant state of the object and $e_{\nu,i}$ is the light curve produced by one event (the index $i$ indicates that the event light curve may be different for each event), and $t_i$ is the ``birth time'' of the $i^{\mathrm{th}}$ event. It is a generalization of a shot-noise process. This representation  can apply to the blue-bump of the \SF, as the variations are quasi-simultaneous from the ultraviolet to the optical (Courvoisier \& Clavel \cite{CC91}). Moreover it is very general, and can be a reasonable approximation of many models, from accretion disk models, e.g.\ the ``blob'' model (Haardt et al.\ \cite{HMG94}) or shocks in the disk (Chakrabarti \& Wiita \cite{CW93}), to supernova explosions (Terlevich \cite{T92}) and relativistic star collisions (Courvoisier et al.\ \cite{CPW96}). The interesting point of the discrete-event models is that the variability properties of a set of object is subject to strong constraints as soon as one puts limitations on Eq.~(\ref{eq-dis_ev}).

To demonstrate analytically these constraints, we develop the following formalism: The mean luminosity density $\overline{L_\nu(t)}$, which can be written (in the following we omit the ``$\nu$'' index, the terms ``luminosity'' and ``energy'' must be understood as ``luminosity density '' and ``energy density'', and we assume that the studied systems are completely stationnary):
\begin{equation}
\label{eq-mean}
\BL=\lim_{T\rightarrow \infty}~ \frac{1}{2T}~ \int_{-T}^{+T} ~L(t)~\dd t=\BN\cdot \overline{E}+C,
\end{equation}
where $\BN$ is the event mean rate, $E_i=\int_{-\infty}^{+\infty} e_i(t-t_i)~\dd t$ is the total energy released by the $i^{\mathrm{th}}$ event (at frequency $\nu$) and $\overline{E}$ is the mean of $E_i$ over all $i$'s. Eq.~(\ref{eq-mean}) is valid even if the $e_i(t)$ differ from each other.  

It is much more complicated to obtain the variance of $L(t)$. Its exact definition is:
\begin{equation}
\label{eq-var_reel}
\Var(L)=\lim_{T\rightarrow \infty}~ \frac{1}{2T}~ \int_{-T}^{+T} ~\left( L(t)- \BL \right)^2~\dd t
\end{equation}
Using the hypothesis that the light curve is given by Eq.~(\ref{eq-dis_ev}), it is possible to transform this equation. The analytical development is made in Appendix A and the result is given in Eq.~(\ref{eq-an_var_all}). If one makes the assumption that the event rate is independent of the parameters defining the events, one finds that (from Eq.~(\ref{eq-an_var_ind})):
\begin{equation}
\label{eq-var}
\Var(L)\sim \BN
\end{equation}
The relative variability, i.e.\ the ratio between the standard deviation (the square root of the variance) and the mean, is therefore:
\begin{equation}
\label{eq-sigma}
\sigma(L)\sim\frac{\sqrt{\BN}}{\BN\cdot \overline{E}+C}
\end{equation}
The strongest limiting case on Eq.~(\ref{eq-dis_ev}) has already been discussed by Pica \& Smith (\cite{PS83}): the whole luminosity is due to the events ($C=0$) and all events are independent and identical. Eq.~(\ref{eq-sigma}) shows that $\sigma(L)$ is proportional to $\BN^{-1/2}$, i.e.\ proportional to $\BL^{-1/2}$ (by Eq.~(\ref{eq-mean})). However, Eq.~(\ref{eq-sigma}) is much more general; in particular the events may be completely arbitrary, and the $-1/2$ index will still be preserved. $C=0$ is not even necessary. The $-1/2$ index is also obtained if $C$ is a fixed fraction of the total luminosity ($C=x\cdot\BL,~ 0< x< 1$). Moreover, if $C$ is constant for all objects (e.g.\ if $C$ is the contribution of the underlying galaxy), we still find the $-1/2$ index as soon as $C$ is negligible compared to the events' luminosity. In conclusion, we see that the only requirements to preserve the index $-1/2$ is that the parameters of the events are drawn from the same distributions in all objects (i.e.\ the distributions of the parameters are independent of $\BN$).

Pica \& Smith (\cite{PS83}) concluded therefore that such models can be rejected independently of the nature of the events (although they did not prove this rigourously). They have found that the $-1/2$ index could be dismissed with a confidence larger than 99 \%. PC94 and Hook et al.\ (\cite{Hal94}) have reached a similar conclusion. However in all these studies the variability have been made in the observer's frame. In this paper, the variability used to determine the relationship with the luminosity is corrected for the redshift effect, and here again the value $-1/2$ can be rejected with a large confidence.

Terlevich (\cite{T92}) proposed that the AGN could be powered by independent supernova explosions. In this model, the events (the supernovae) are identical and independent, and $C=x\cdot\BL,~ 0\le x< 1$ is satisfied, because a fixed fraction of the total luminosity is emitted by normal (constant) stars ($x$ is larger than 0, but depends strongly, of course, on the considered wavelength). As we have seen, the index of the relationship between luminosity and variability remains equal to $-1/2$.

Even if a general trend is predicted, we see that much room is left for the observed intrinsic scatter in the terms inside the integrals in Eq.~(\ref{eq-an_var_ind}). For instance the distributions of the luminosities of the events may differ from object to object without being necessarily related to the total luminosity of the object.

As the observations clearly prohibit the so universal $-1/2$ index, we could ask whether the release of one of the two properties of the events (``identity of the distributions of the event parameters'' and ``independence'') might produce the observed behaviour.

\subsection{Non-identity of the distributions of the event parameters\label{nonid}}

The variability--luminosity relationship can be modified to match any given relationship if one adjusts in an {\em ad-hoc} way the distributions of the event parameters. We examine only two very simple cases that can be intuitively expected in many discrete-event models.

\subsubsection{Scaling of the event luminosity}
A very simple modification that breaks the ``identity of the distributions of the event parameters'' assumption is that the energy released by the events scales with the total luminosity of the object, while the events are still independent from each other. We can estimate the effect of such a scaling on the variability--luminosity relationship using the notation of Eq.~(\ref{eq-scal_not}) and the results of Eqs~(\ref{eq-mean_scal}) and (\ref{eq-var_scal}) (see Appendix B) in the case $C=x\cdot \BL$,~ $0\le x <1$. The scaling is written $\overline{E}=k\cdot \overline{E_0}\sim \BL^\alpha$, which gives $\BN\sim \BL^{1-\alpha}$. Therefore Eq.~(\ref{eq-sigma}) becomes:
\begin{equation}
\label{eq-scal_lum}
\sigma(L)\sim\frac{\sqrt{\BL^{1-\alpha}\cdot \BL^{2\alpha}}}{\BL}\sim \BL^{\;(\alpha-1)/2}
\end{equation}
If no correlation exists between variability and luminosity ($\eta=0$), then we must have: $\alpha=1$. If the $\eta=-0.08$ index is correct, one obtains:
\begin{equation}
\alpha\simeq 0.84
\end{equation}
This scaling cannot be applied to models based on supernovae, as the energy released by a supernova is essentially fixed. On the other hand, it is reasonable to assume that the luminosity of an event -- of unspecified nature -- related to the existence of a black hole scales -- in a way or in another -- with its mass. For instance, Courvoisier et al.\ (\cite{CPW96}) have proposed a model where such a scaling with luminosity could be rather natural. In this model, the events are collisions between stars in a dense stellar cluster surrounding a supermassive black hole, which liberates the stars' relativistic kinetic energy. For a given cluster, an increase of the black-hole mass will increase the energy liberated by the events.

Generally speaking, however, the value of $\alpha$ shows that the event luminosity is not trivially related to the total luminosity. This may be a major difficulty for the discrete-event models.

\subsubsection{Scaling of the event stretching\label{stretch}}
We call ``life time'' of an event any time that is mathematically associated with the event. For instance, the life time of a ``box'' event is the duration of the non-zero state; for an semi-exponential event, the life time may be the $e$-folding time. We may think that the event life time increases with the total luminosity. This may appear naturally in many models. For instance, in case of reprocessing on an accretion disk, the size of the reprocessing area (contributing at a given frequency) very certainly increases with the luminosity of the X-ray source(s). We call the possibility to modify the event life time the ``stretching'' of the events.

We introduce in the mathematical formulation of the events a stretching factor $\ell$ (corresponding to ``$p_2$'' in Eq.~(\ref{eq-scal_not})). We assume that $\ell$ scales with the total luminosity of the object, i.e.\ $\overline{\ell}=\delta\cdot \overline{\ell_0}\sim \BL^\beta$ (see Appendix B). From Eqs~(\ref{eq-mean_scal}) and (\ref{eq-var_scal}) we obtain (valid if $C=x\cdot\BL,~ 0\leq x < 1$):
\begin{equation}
\label{eq-scal_dur}
\sigma(L)\sim \frac{\sqrt{\BL\cdot \BL^{-\beta}}}{\BL}\sim \BL^{\;-(\beta+1)/2}
\end{equation}
If $\eta=0$, we must have $\beta=-1$; with our value for $\eta$ we find:
\begin{equation}
\beta\simeq -0.84
\end{equation}
These values are not satisfactory, because one would rather expect that the stretching factor increases with the luminosity, and not decreases. It appears therefore that this mechanism (alone, at least) cannot explain the variability--luminosity relationship.

\subsubsection{Scaling of both the event luminosity and the event stretching}
We can also combine the scaling of the event luminosity and the one of the event stretching. Again, the justification of these scalings is that they seem natural to us. Eqs~(\ref{eq-mean_scal}) and (\ref{eq-var_scal}) give here:
\begin{equation}
\label{eq-scal_both}
\sigma(L)\sim \frac{\sqrt{\BL^{1-\alpha}\cdot \BL^{2\alpha}\cdot \BL^{-\beta}}}{\BL}\sim \BL^{\;-1/2+(\alpha-\beta)/2}
\end{equation}
If we assume that $\beta=1$, which means that the stretching factor of the events is proportional to the total luminosity, our result requires:
\begin{equation}
\alpha\simeq 1.84
\end{equation}
or, if $\alpha=1$, i.e.\ the event luminosity is proportional to the total luminosity:
\begin{equation}
\beta\simeq 1.84
\end{equation}
If we impose $\alpha\geq 0$ and $\beta\geq 0$, which is physically the most appropriate, we find that $\alpha$ must be $\geq 0.84$, because $\beta\simeq \alpha-0.84$. If $\eta=0$, we must have $\alpha\geq 1$, and $\beta=\alpha-1$. The models are at the moment not sufficiently developed to predict any value for $\alpha$ and $\beta$. We can however remark that the constraints on $\alpha$ and $\beta$ are in no case intuitively ``natural''.

\subsection{\label{depend}Interdependence of the events}
\subsubsection{Pseudo-Poisson process}
It has often been argued that a different relationship between variability and luminosity could be obtained if the events were not independent from each other.
For instance, one can imagine a cluster of explosive charges. The charges may explode spontaneously (producing an ``event''), or the explosion may be triggered by the close explosion of another charge. It is easy to imagine a mathematically analogous situation in AGN.

In this picture, each spontaneous event (not triggered by another event) is followed by a sequence of ``child'' events. Therefore one can replace the spontaneous event and its ``child'' events by one ``super-event''. The normalization and shape of this ``super-event'' are not constant, but are drawn from complex distributions.

It appears therefore from the discussion in Sect.~\ref{proc_Poiss} that the $-1/2$ index cannot be avoided by any process following this picture. For instance, if all the events are independent and if the ``parent'' event has a probability $\varphi$ to produce a simultaneous ``child'' event, which in turn has a probability $\varphi$ to produce another ``child'' event, and so on, the variability at a given total luminosity (if $C=0$) is multiplied by: $\sqrt{1+\varphi}/\sqrt{1-\varphi}$.

\subsubsection{Non-Poisson process}
The reason why the $-1/2$ index always holds in the processes dicussed above is that the process remains Poissonian in essence. However no clear physical picture of non-Poisson process can be proposed at the moment. We can nevertheless propose a mathematical non-Poisson process:

In a system where events occur according to the Poisson law, the probability $P_0(t_1)$ that no event occurs between $t=0$ and $t=t_1$ follows the equation:
\begin{equation}
\label{eq-Poiss}
\frac{\dd P_0}{\dd t}= -\mu\cdot P_0,
\end{equation}
with the boundary condition $P_0(t=0)=1$, $\mu$ being the mean number of events between $t=0$ and $t=1$. The solution of Eq.~(\ref{eq-Poiss}) is simply:
\begin{equation}
P_0(t)= e^{-\mu t}
\end{equation}

We can modify Eq.~(\ref{eq-Poiss}), so that the mean time without event decreases when new events occur; this reflects the interdependence of the events. We write:
\begin{equation}
\label{eq-non_Poiss}
\frac{\dd P_0}{\dd t}= -\left(\mu+\varrho\cdot\sum_{k=1}^{n}~f(t-t_k)\right)\cdot P_0,
\end{equation}
where $\varrho$ is the ``response'' to the occurence of an event --  whatever its origin --, $n$ is the number of events that have already happened, and $f(t)$ is the time distribution of the influence of an event, with $\int_0^\infty~f(t)~\dd t=1$. $\varrho$ should be smaller than 1, as a larger value would produce a diverging ``chain reaction''.

\begin{figure}[p]
\centerline{\mbox{\psfig{file=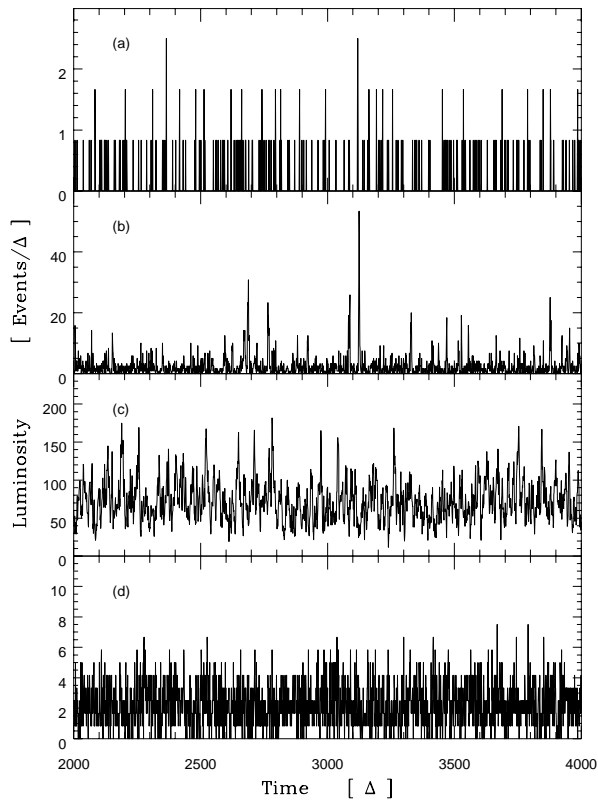,width=8.4cm}}}
\caption{\label{depclum}{\bf a-c.} Portion of light curves simulated with $\varrho=0.9$ in the three different regimes: {\bf a} sub-critical, $\BL=0.1$ events/$\Delta$, {\bf b} critical, $\BL=2.2$ events/$\Delta$, {\bf c} super-critical, $\BL=72$ events/$\Delta$. {\bf d} Light curve simulated with $\varrho=0$, and the same mean luminosity as in panel b}
\end{figure}

It is possible to solve Eq.~(\ref{eq-non_Poiss}) numerically for each new event
and to draw lists of successive events for a given choice of $\mu$ and $\varrho$. If we choose an event light curve, we can construct the total light curve, and deduce the luminosity and the variability of the virtual object. In all the simulations, the distribution function $f(t)$ has been taken to be uniform between 0 and 1 (arbitrary) unit of time ($\Delta$) and the event light curve was a simple box function. For a set of $\varrho$ chosen between 0 and 0.96, we calculated different light curves, making $\mu$, the rate of independent events, vary from $10^{-3}$ to as many events per $\Delta$ as possible (the limitation is due to the calculation time, which increases with the rate of events). Examples of light curves are shown in Fig.~\ref{depclum}.

\begin{figure}[p]
\centerline{\mbox{\psfig{file=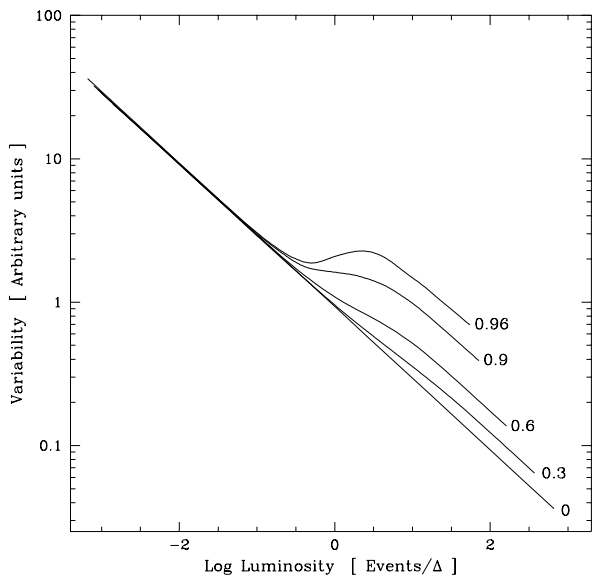,width=8.4cm}}}
\caption{\label{simvarlum}Variability as a function of the luminosity (expressed in number of events per $\Delta$) in the simulation of interdependent events for different values of $\varrho$. The $\varrho=0$ curve has a $-1/2$ slope.}
\end{figure}
Figure~\ref{simvarlum} shows the variability--luminosity relationship obtained from the simulations for different values of $\varrho$. In all curves the behaviour is comparable; one can distinguish 3 regimes:
\begin{itemize}
\item The {\em sub-critical regime} occurs when the mean rate of events is small. The variability--luminosity relationship has the slope predicted by pure Poisson statistics, independently of $\varrho$. The events are too rare to affect significantly Eq.~(\ref{eq-non_Poiss}). The light curves (Fig.~\ref{depclum}a) do not show clear structures (as is expected from pure Poisson process).
\item The {\em critical regime} occurs when the mean rate of events is around 1 per $\Delta$ (slightly dependent on $\varrho$). The variability decreases less rapidly, or even increases for large values of $\varrho$. The larger $\varrho$, the larger the deviation from the $-1/2$ index. In this regime, the event density is right in the range that produces ``chain reaction''. The light curves  show very characteristic, flare-like structures due to the interdependence of the events (e.g.\ around $t=3100$ $\Delta$ in Fig.~\ref{depclum}b).
\item The {\em super-critical regime} occurs when the mean rate of events increases further. The variability--luminosity relationship has again the slope predicted by pure Poisson statistics, but with a larger normalization than in the sub-critical regime. It seems that the ``chain reaction'' is saturated by the large event density, and the light curves (Fig.~\ref{depclum}c) have essentially lost the characteristic structures, probably because of the ``overcrowding'' of the events. The increase of variability in this regime as a function of $\rho$ is compatible with the multiplication by a factor $\sqrt{1+\varrho}/\sqrt{1-\varrho}$.
\end{itemize}

Even though this model can produce a variability--luminosity relationship different from the Poisson case, none of the curves in Fig.~\ref {simvarlum} can be adjusted to the data in Fig.~\ref{Blum}. Moreover, the physical feasability of this model cannot be asserted. Other mathematical formulations might yield more appropriate results. However, the physical feasability might be a major difficulty in most (if not all) of them, because (pseudo-)Poisson processes appear much more natural.

\section{Conclusion}
The trend between the variability and the luminosity seen by many authors is confirmed when one estimates the rest-frame variability. The most probable index of the relationship is $-0.08$. It is compatible with results obtained using high-luminosity quasars. This is an additional argument for the continuity between the Seyfert galaxies and the most powerful quasars.

We have investigated the effects of several biases, and we consider our result as very probably correct. An intrinsic $\eta=-1/2$ index, which is what is generally expected from discrete-event models, would be possible only if there exists very large and yet undiscovered biases; but this seems highly unlikely. The other extreme, which is that the variability does not depend on the luminosity, is more plausible. However the difficulty in producing such a relationship with discrete events is not significantly alleviated. 

The easiest way to produce a relationship with an index different from $-1/2$ is to assume that the average event luminosity increases with luminosity. In this case, it would mean that the event luminosity must increase only about 15 \% more slowly than the total luminosity. The scaling of the event stretching factor may also contribute to the relationship. The relationships given here may provide useful constraints for any kind of discrete-event model.

The possibility that the birth time of the events does not follow a Poisson distribution has also been investigated and cannot be completely excluded. However the system that we have explored cannot (alone) produce the correct relationship. Anyway a non-Poisson system appears somehow unnatural, and the physical situation at its origin might be difficult to find.

The predictions of physical models that can be expressed in terms of discrete events can be compared quantitatively with the variability--luminosity relationship using Eq.~(\ref{eq-an_var_ind}). But, it appears to us that discrete-event models do not satisfy the variability--luminosity relationship in a natural way. This difficulty possibly means that this kind of model does not apply to the active galactic nuclei, and, therefore, that Eq.~(\ref{eq-dis_ev}) is not valid. Other source of variability (not considering the possibility that variations are extrinsic, which seems very unlikely) could be global effects, which do not have a simple general analytical formulation, in opposition to the local processes discussed here. 

\begin{acknowledgements}
The careful and critical reading of the referee has contributed to improve significantly the mathematical aspects of this paper. SP gratefully thanks the system managers of the Geneva Observatory and of the ISDC for having allowed him to run the simulations. SP also aknowledges a grant from the Swiss National Science Foundation.
\end{acknowledgements}

\appendix\section{Appendix: Derivation of the variance of $L(t)$}
The variance of the generalized shot-noise process given in Eq.~(\ref{eq-dis_ev}) can be calculated using Campbell's theorem (e.g.\ Papoulis \cite{P91}), which states that if:
\begin{equation}
s(t)= \sum_i~ h(t-t_i),
\end{equation}
where the points $t_i$ follow a Poisson distribution with a mean rate $\lambda$, then the variance of $s(t)$ is given by:
\begin{equation}
\Var(s)= \lambda\cdot \int_{-\infty}^{+\infty}~ h^2(t)~\dd t
\end{equation}

In our case, everything happens as if we have a superimposition of an infinite (possibly continuous) number of shot noises given by the relationship:
\begin{equation}
L_{\textstyle\vec{p}}= \sum_{i_{\textstyle\vec{p}}}~ e_{\textstyle\vec{p}}(t-t_{i_{\textstyle\vec{p}}})
\end{equation}
(We neglect the constant term, because it does not appear in the variance). The $\vec{p}$ vector is the list of parameters that define the event; $\vec{p}$ is an element of the $\cal D(\vec{p})$ space, of arbitrary dimension, and follows a distribution $P(\vec{p})$ with $\int_{\cal D(\textstyle\vec{p})} \dd P(\vec{p})= 1$. The average rate of these shot noises are $\BN_{\textstyle\vec{p}}\cdot \dd P(\vec{p})$. For each $\vec{p}$, the variance of the shot noise is:
\begin{equation}
\dd\Var(L_{\textstyle\vec{p}})=\BN_{\textstyle\vec{p}}\cdot \dd P(\vec{p})\cdot \int_{-\infty}^{+\infty} e^2_{\textstyle\vec{p}}(t)~ \dd t
\end{equation}
Therefore the variance of the total light curve is the sum (integration) of the variance of all shot noises:
\[
\Var(L)=\int_{\cal D({\textstyle\vec{p}})} \dd \Var(L_{\textstyle\vec{p}})=
\]
\begin{equation}
\label{eq-an_var_all}
\phantom{\Var(L)}=\int_{\cal D({\textstyle\vec{p}})} \BN_{\textstyle\vec{p}}\cdot \int_{-\infty}^{+\infty} e^2_{\textstyle\vec{p}}(t)~ \dd t~ \dd P(\vec{p})
\end{equation}

If one make the assumption that the rates $\BN_{\textstyle\vec{p}}$ are independent of $\vec{p}$, Eq.~(\ref{eq-an_var_all}) becomes: 
\begin{equation}
\label{eq-an_var_ind}
\Var(L)=\BN\cdot \int_{\cal D({\textstyle\vec{p}})} \int_{-\infty}^{+\infty} e^2_{\textstyle\vec{p}}(t)~ \dd t~ \dd P(\vec{p}),
\end{equation}
which gives Eq.~(\ref{eq-var}).

\section{Appendix: Effects of the normalization and of the stretching factor on $\sigma(L)$}
Eq.~(\ref{eq-an_var_all}) can be developed further if we assign physical meanings to some components of $\vec{p}$. The normalization (i.e.\ the total energy liberated by an event) and the stretching factor are respectively the first and the second components of $\vec{p}$. We can write:
\begin{equation}
\label{eq-scal_not}
e_{\textstyle\vec{p}}(t)=p_1\cdot \frac{1}{p_2}\cdot e_{\textstyle\hat{\vec{p}}}\left(\frac{t}{p_2}\right),
\end{equation}
where $\hat{\vec{p}}=(p_3,\ldots)$. We impose $\int_{-\infty}^{+\infty}~ e_{\textstyle\hat{\vec{p}}}(t)~ \dd t=1$, so that we have $E_{\textstyle\vec{p}} \equiv \int_{-\infty}^{+\infty}~ e_{\textstyle\vec{p}}(t)~ \dd t=p_1,~ \forall \vec{p}$. The reason why $e_{\textstyle\hat{\vec{p}}} (\frac{t}{p_2})$ is multiplied by $p_2^{-1}$ is that it makes $\int_{-\infty}^{+\infty}~ e_{\textstyle\vec{p}}(t)~\dd t$ be independent of $p_2$ (because $\int_{-\infty}^{+\infty}~ f(k\cdot x)~\dd x=\frac{1}{k}\cdot \int_{-\infty}^{+\infty}~f(x)~\dd x$). Let us make the additional assumption that $p_1$ and $p_2$ are independent of each other and of the other $p_i$'s. Let us write $\cal D(p_1)$, $\cal D(p_2)$, and $\cal D(\hat{\vec{p}})$ the spaces from which are drawn $p_1$, $p_2$, and $\hat{\vec{p}}$ respectively. This makes Eq.~(\ref{eq-an_var_ind}) become:
\[
\Var(L)=\BN\cdot \int_{\cal D({\textstyle\vec{p}})} \int_{-\infty}^{+\infty} p_1^2\cdot \frac{1}{p_2^2} \cdot e^2_{\textstyle\hat{\vec{p}}}\left(\frac{t}{p_2}\right)~ \dd t~ \dd P(\vec{p})=
\]
\[
\phantom{\Var(L)}=\BN\cdot \int_{\cal D(p_1)} \int_{\cal D(p_2)} \int_{\cal D(\textstyle\hat{\vec{p}})} \int_{-\infty}^{+\infty} p_1^2\cdot \frac{1}{p_2^2} \cdot
\]
\[
\hspace*{92pt} e^2_{\textstyle\hat{\vec{p}}}\left(\frac{t}{p_2}\right) \dd t~ \dd P(\hat{\vec{p}})~ \dd P(p_2)~ \dd P(p_1)=
\]
\[
\phantom{\Var(L)}=\BN\cdot \overline{p_1^2}\cdot \int_{\cal D(p_2)} \frac{1}{p_2^2}~ \int_{\cal D(\textstyle\hat{\vec{p}})} \int_{-\infty}^{+\infty}
\]
\[
\hspace*{122.7pt}e^2_{\textstyle\hat{\vec{p}}}\left(\frac{t}{p_2}\right)~ \dd t~ \dd P(\hat{\vec{p}})~ \dd P(p_2)=
\]
\[
\phantom{\Var(L)}=\BN\cdot \overline{p_1^2}\cdot \int_{\cal D(p_2)} \frac{1}{p_2}\cdot \int_{\cal D(\textstyle\hat{\vec{p}})}\int_{-\infty}^{+\infty}
\]
\[
\hspace*{141.3pt}e^2_{\textstyle\hat{\vec{p}}}(t)~ \dd t~ \dd P(\hat{\vec{p}})~ \dd P(p_2) =
\]
\begin{equation}
\label{eq-an_var_restr}
\phantom{\Var(L)}=\BN\cdot \overline{p_1^2}\cdot \overline{p_2^{-1}}\cdot
\left(\int_{\cal D(\textstyle\hat{\vec{p}})} \int_{-\infty}^{+\infty} e^2_{\textstyle\hat{\vec{p}}}(t)~ \dd t~ \dd P(\hat{\vec{p}}) \right)
\end{equation}
This relationship is used to derived the scaling properties in Sect.~\ref{nonid}. To obtain the results shown in this section, we introduce two random variables, $E_0$ and $\ell_0$, with the relations: $p_1\equiv E=k\cdot E_0$ and $p_2\equiv \ell=\delta\cdot \ell_0$ (we call $K$ the multiple integral). The above equation becomes:
\begin{equation}
\label{eq-var_scal}
\Var(L)=\BN\cdot \left(k^2\cdot \overline{E_0^2}\right)\cdot \left(\delta^{-1}\cdot \overline{\ell_0^{-1}}\right)\cdot K
\end{equation}
With the same substitutions, the mean luminosity becomes:
\begin{equation}
\label{eq-mean_scal}
\BL=\BN\cdot \left(k\cdot \overline{E_0}\right) + C
\end{equation}
Eqs.~(\ref{eq-mean_scal}) and (\ref{eq-var_scal}) give directly Eqs.~(\ref{eq-scal_lum}), (\ref{eq-scal_dur}), and (\ref{eq-scal_both}).

\end{document}